\documentclass[12pt]{article}
\usepackage{makeidx}         % allows index generation
\usepackage{graphicx}        % standard LaTeX graphics tool
\usepackage{multicol}        % used for the two-column index
\usepackage[bottom]{footmisc}% places footnotes at page bottom

\usepackage{graphicx}
\usepackage{amsfonts,amssymb,amsmath,paralist,wasysym,cancel}
\usepackage{epsfig,rotating,framed,enumerate,listings,subfigure}

\usepackage{wrapfig,nicefrac}
\usepackage{marginnote}
\usepackage{booktabs}

\usepackage{units}
\usepackage[raggedright]{sidecap}
\usepackage{algorithm}
\usepackage[noend]{algpseudocode}

\newcommand{\used}{\text{\it used}}

\newcommand{\von}{\text{\it from}}
\newcommand{\bis}{\text{\it to}}

\makeatletter
\renewcommand{\@fnsymbol}[1]{\@arabic{#1}}
\makeatother

\begin{document}

\title{Column-generation for a two-dimensional multi-criteria bin-packing problem}

\author{Christof Groschke, e-mail: {\small\tt christof.groschke@stud.hn.de}
\and Steffen Goebbels\thanks{corresponding author: iPattern Institute, Niederrhein University of Applied Sciences, Reinarzstr.~47, 47805 Krefeld, Germany}, e-mail: {\small\tt steffen.goebbels@hsnr.de}%\orcidID{0000-0003-4313-9101}
\and Jochen Rethmann, e-mail: {\small\tt jochen.rethmann@hsnr.de}}
\date{}
\maketitle

\begin{abstract}
In this study, we examine a two-dimensional bin-packing problem in printed circuit board
manufacturing. 
Among other objectives, the number of bins, but also the number of different bin layouts, is 
to be minimized. As the running times of an earlier MIP presentation are only
acceptable for small problem instances, we will now discuss a
branch-and-price approach 
by using an adapted Ryan-Foster-bran\-ching. 
The pricing problem computes the layouts, separating the 
time-consuming constraints from the master problem.

\medskip \noindent
{\bf Key words} 2D Bin-Packing; Mixed Integer Linear Program, Column-Generation, Branch-and-Price
\end{abstract}

%%%%%%%%%%%%%%%%%%%%%%%%%%%%%%%%%%%%%%%%%%%%%%%%%%%%%%%%%%%%%%%%%%%%%%%%%%
\section{Introduction}
%%%%%%%%%%%%%%%%%%%%%%%%%%%%%%%%%%%%%%%%%%%%%%%%%%%%%%%%%%%%%%%%%%%%%%%%%%
In the production of printed circuit boards, multiple copies (items) of smaller rectangular board
types ($n$ item types) must be placed within larger boards (rectangular bins) for production 
while maintaining certain 
spacing constraints between the items, depending on whether item edges need to be carved or milled. 
In the classical
2D bin-packing problem, the number of bins has to be minimized. In addition, in the given problem
the number of bin patterns (the arrangements or layouts of items within a bin) should also be minimized to reduce the manual processing effort.
This is supported by allowing the number of items of the same type $j$ to vary within a given range
$\von_j$ to $\bis_j$. 
A corresponding MIP is presented in \cite{or2024}. To avoid a non-convex
quadratic problem, the MIP uses multiple bins for each pattern that are either empty or containing
exactly that pattern. Another disadvantage is that patterns with equal item count are different
if positions of items are not the same -- leading to an exploding number of different patterns.
To eliminate these disadvantages, we decouple the arrangement of items within patterns from a 
master problem that searches for a minimum number of patterns and bins. 
This idea is applied in \cite{Pisinger07} to solve the 2D bin-packing problem.

We represent a pattern $l$ by a layout vector $\vec{a}_l=(a_{1,l},\dots a_{n,l})$, where
coordinate $a_{j,l}$ is the number of items of type $j$ in that pattern. This representation
is independent of the actual positions of individual items in bins as well as of their 
assignment to individual bins. This allows many of the symmetric solutions of the
MIP in \cite{or2024} to be avoided.
Let $x_l\in\mathbb{N}_0:=\{0, 1, 2,\dots\}$ be the number of bins that are filled with 
pattern $l$.
If all $p$ feasible patterns can be enumerated, a simplified version of this 
bin-packing problem is
\begin{alignat}{1}
\text{minimize } & \sum_{l=1}^p (c_1\cdot \used_l + c_2\cdot x_l)\label{eqMIPMaster}\\
\text{subject to }& \frac{1}{M}\cdot x_l\leq \used_l \text{ for all } l\in[p]:=\{1, 2,\dots,p\},\label{eqConst}\\
& s_j = \sum_{l=1}^p a_{j,l}\cdot x_l,~
 \von_j \leq s_j \leq  \bis_j\text{ for all } j\in[n],\label{eqMIPRange}
\end{alignat}
where $M$ is an upper bound for the number of items of any specific type. The binary
variables $\used_l\in\{0, 1\}$ indicate whether the pattern $l$ is used, i.e., whether $x_l>0$.
Constant $c_1>0$ weights the goal to minimize the number of different patterns, constant $c_2>0$
weights the objective to minimize the number of bins.

The idea of column-generation is to start with only a subset of $p'$ patterns (initially
generated by a simplified version of the MIP in \cite{or2024}) and interpret
the MIP (\ref{eqMIPMaster}--\ref{eqMIPRange}) as a restricted master problem by replacing $p$ with $p'$. Its LP-relaxation is solved to optimality to obtain
values of an optimal solution of the dual problem (shadow prices). With these values, the next pattern (column) $p'+1$ can be computed
by solving a pricing problem. This is a MIP that especially considers distance constraints.
Then the new restricted master problem with $p'+1$ columns (patterns) is solved, and so on. 
The procedure ends when no
new pattern can be generated that improves the objective value of the relaxed master problem.  

Unfortunately, relaxation of the problem eliminates the objective to minimize the number of patters.
If $\used_l$ is a float variable, then $\used_l=\frac{1}{M}\cdot x_l$ because of (\ref{eqConst}), and
$$ \sum_{l=1}^p (c_1\cdot \used_l + c_2\cdot x_l)
= \sum_{l=1}^p \left(\frac{c_1}{M} +c_2\right) \cdot x_l.
$$
Thus, only the goal to minimize the number of bins remains, and we can replace the constant
$\frac{c_1}{M} +c_2$ by 1. To address this problem and to
cope with non-integer values of the relaxed, restricted master problem in standard form
\begin{alignat*}{2}
\text{maximize }  \sum_{l=1}^{p'}  (-1) \cdot x_l
\text{ subject to } 
& \sum_{l=1}^{p'} -a_{j,l}\cdot x_l \leq -\von_j, &\qquad (\text{dual } \pi_{j,1})\\
& \sum_{l=1}^{p'} a_{j,l}\cdot x_l \leq \bis_j,  &\qquad  (\text{dual } \pi_{j,2})
\end{alignat*}
for all $j\in[n]$, we use a branch-and-price strategy
based on Ryan-Foster-branching with a new %problem specific
selection rule that favors small numbers of patterns. 

\section{Branch-and-Price}
The pricing problem 
is to find the next column $(a_{1,p'+1},\dots,a_{n,p'+1})^\top$ to be
considered. This is done based on the shadow prices $\pi_{j,1}$, $\pi_{j,2}$ obtained by solving the 
relaxed, restricted master problem 
via the MIP
\begin{equation}
 \text{maximize } -1 + \sum_{j=1}^n (\pi_{j,1}-\pi_{j,2}) \cdot a_{j,p'+1}\label{eqPrice}
\end{equation}
subject to $a_{j,p'+1}\in\{0,\dots,\bis_j\}$, $j\in[n]$, and the constraints that the items of all types $j$
can be placed in one bin with the given multiplicities $a_{j,p'+1}$ while avoiding
overlap and maintaining distance constraints.
If the maximum is not positive, then the maximum previously found using the relaxed, 
restricted master problem 
is also the maximum of the relaxed, unrestricted master problem. Otherwise, a new column 
that is different from the first $p'$
columns has been found. This column has a maximum reduced cost.

In order to address the runtime performance issues associated with the MIP of the pricing 
problem, we employ a straightforward greedy approach involving the iterative 
increase of item multiplicities and the verification of patterns using a bottom-left 
placement algorithm.
Based on the ideas described in \cite[p. 44]{Pisinger07}, different sequences of item types are 
generated and then run through in parallel.
One sequence results from sorting the item types in descending order based on values $\pi_{j,1}-\pi_{j,2}$.
Another sequence results from sorting these numbers divided by item areas. Instead of strictly 
sorting, sequences are also generated randomly by interpreting these values as probabilities.
Iteratively for each
item type in a sequence, the algorithm increases its item multiplicity from 0 as long as 
the items fit into the pattern according to a bottom-left placement.
From the results of all sequences, all columns with a positive value of the objective function in 
(\ref{eqPrice}) are added to the restricted master problem that then can have even more than $p'+1$ columns.

Due to the relaxation of the master problem, the column-generation approach
may produce non-integral solutions. Therefore, a branch-and-bound algorithm -- known as 
a branch-and-price algorithm in the 
context of column-generation -- is required to find
integer values for the variables $x_l$. 
At each node of the branch-and-bound tree, the column-generation algorithm is used to solve a relaxed 
master problem. To obtain integer values, one could branch on master variables such as $x_l$. 
In general, ``most fractional branching'' is not recommended, see \cite{Achterberg05}.
A favorable alternative 
appears to be the branching rule developed by Ryan and Foster (cf.~\cite[p. 474]{Luebbecke24}). 
When applied to a binary 1D bin-packing problem, in which exactly one item of each type must be placed, 
the rule splits the problem into two sub-problems by considering two item types. In the first sub-problem, the two 
corresponding items must be placed in the same bin. This can be achieved by replacing the two item types with 
a compound version. 
In the second sub-problem, the items of the two item types must not be in the same bin. This can be achieved
through additional constraints in the pricing problem.
We use the solution described in \cite{daSilva:24} to adapt this strategy to our problem dealing 
with multiple items of one type. In the second sub-problem, items of the two selected types must 
not be in the same bin. In the first sub-problem, however, there must be at least one bin 
containing items of both types.
To this end, we select a pair of items to branch on by calculating the ``affinity'' for each pair 
$(i, j)$ of item types in a column-generation solution with $p'$ columns (patterns): 
\begin{alignat*}{1}
\rho_{i, j} := 
\begin{cases}
  \sum_{l=1}^{p'}  a_{i,l} \cdot \frac{a_{i,l} -1}{2} \cdot x_l, & \text{ if } i = j,\\
  \sum_{l=1}^{p'}  a_{i,l} \cdot a_{j,l} \cdot x_l, \quad  & \text{ otherwise.}
\end{cases} %\label{eq:affinity}
\end{alignat*}
The affinity indicates how frequently items of types $i$ and $j$ are placed together in bins.
For binary bin-packing problems, it is shown in \cite{daSilva:24} that $\rho_{i, j} \in \mathbb{Z}$, 
$i, j\in[n]$, implies that the $x_l$ are integers. Although the authors demonstrated 
that integral affinities do not always imply integral solutions for non-binary problems, we 
select an item type pair $(i, j)$ 
with a maximum value $|\rho_{i, j}-\operatorname{round}(\rho_{i, j})|>0$ for branching
(including pairs where $i=j$) 
if possible. If the affinity of all pairs of 
item types is integral but a variable is still fractional, then we
first select a pattern $l$ 
from the set of all patterns containing 
items of two or more item types 
or of one item type with $\bis_j>1$ such that $x_l - \lfloor x_l \rfloor$ is maximal. 
Even in this case, we do not branch on $x_l$. 
Instead, we determine the item type $i$ whose items cover a largest area within the selected pattern.
If $\bis_i>1$, we select the pair $(i, i)$. Otherwise, we determine an other item type, $j$, whose items
cover the next largest area in the pattern and proceed with the pair $(i, j)$. 
Based on the selected pair $(i, j)$, we create a left and a right child node. 
In the problem of the right node, items of $i$ and $j$ must not be placed in the same bin 
if $i\neq j$.
In the case that $i=j$, at most one item of $i$ is allowed in a bin. To this end and in contrast to
the area condition in \cite{daSilva:24},
explicit constraints are 
added to the corresponding pricing problem.  
The left node problem
is complementary in the sense, that there must be at least one bin containing items of
both $i$
and $j$ (or two items of $i$ if $i=j$). This is implemented by generating a
compound item type (with multiplicities $\von$ and $\bis$ set to one) representing one item 
of $i$ and $j$, respectively, and by decreasing the multiplicities
$\bis_i$, $\bis_j$ and, if still positive, $\von_i$, $\von_j$ by one. This compound item type is 
used in the same way 
as other item types in the master problem. Note that the item types aggregated by compound item types can also 
be compound item types themselves.
The bottom-left heuristic in the pricing problem splits items of compound item types up into 
single items of the original problem
because merging items to obtain a rectangular shape (similar to the branching strategy for a 1D problem) 
may not be possible.
When using the branching heuristic on an item type pair $(i, j)$ again, there may already exist a
compound item type for that pair. In this case, rather than creating a new compound item type, the 
multiplicities ($\von$, $\bis$) of the already existing one are increased.
The heuristic leads to child problems in which a compound item may belong to only one column 
(pattern) $l$. And due to the size of compound items, the columns become unit vectors. 
Then the number of bins belonging to this column, $x_l$,  
must be an integer since its multiplicity, $\von=\bis$, is also an integer and columns are unique.
The problem of a child node inherits the columns of the parent node. However, these columns 
must be 
adjusted according to the branching rule. For a new compound item type, a column representing 
one corresponding item is added. This
additional column is only feasible, if the items that the compound item type represents fit into a bin. 
This is verified using the bottom-left algorithm. If the check fails, 
the corresponding node of the branch-and-bound tree can be discarded. Otherwise, 
the relaxed, restricted master problem will find an optimal solution 
for this set of columns.

Finally, to reduce the number of patterns, we take advantage of the fact that a typical 
branch-and-bound tree contains many solutions with the same number 
of used bins but a different number of required patterns. A pattern $l$ is required if and 
only if $x_l\neq 0$. We apply a simple heuristic selection rule to traverse the tree. Before solving their 
optimization problem, created child nodes are placed in a min-heap, where the key is 
the number of columns 
used in the solution belonging to their parent node. The next problem solved with column 
generation belongs to the node with the smallest key. The node is removed from the 
min-heap, its problem is solved, and its children are added to the min-heap.

\section{Results and Conclusions}
The results of the new selection rule were generated using data from the company. A subset is shown 
in Table \ref{tab:dataRecords}. Each item type $j$ has a width $w_j$ and height $h_j$, given in millimeters, and
must to be produced 
at least $\von_j$ times. 
An overproduction of 15\% is permitted, giving $\bis_j$. 
Each item type needs to be milled. 
Therefore, a distance of 6 mm must be maintained between all items, see \cite{or2024}.
\begin{table}[tb]
\caption{Data records: bin size is $614 \ \mathrm{mm} \times 512 \ \mathrm{mm}$}
\label{tab:dataRecords}\tiny\begin{center}%
\begin{tabular}{r@{\hspace{0.4cm}}rrr@{\hspace{0.25cm}}rrr@{\hspace{0.25cm} }rrr@{\hspace{0.25cm}}rrr@{\hspace{0.25cm}}rrr@{\hspace{0.25cm}}}\toprule
item & \multicolumn{3}{c}{$r_1$}& \multicolumn{3}{c}{$r_2$} & \multicolumn{3}{c}{$r_3$} & \multicolumn{3}{c}{$r_4$} & \multicolumn{3}{c}{$r_5$}\\ 
\cmidrule(lr){2-4} \cmidrule(lr){5-7} \cmidrule(lr){8-10} \cmidrule(lr){11-13} \cmidrule(lr){14-16}
type & $w_j$ & $h_j$ & $\scriptstyle{\von_j}$ & $w_j$ & $h_j$ & $\scriptstyle{\von_j}$ & $w_j$ & $h_j$ & $\scriptstyle{\von_j}$&  $w_j$ & $h_j$ & $\scriptstyle{\von_j}$ & $w_j$ & $h_j$ & $\scriptstyle{\von_j}$ \\ \midrule
1 & 55 & 111 & 50 & 43 & 44 & 10 & 60 & 70 & 3 & 66 & 130 & 6 & 140 & 194 & 25 \\
2 & 73 & 132 & 106 & 194 & 283 & 15 & 93 & 208 & 13 & 210 & 165 & 10 & 8 & 82 & 40 \\
3 & 28 & 55 & 2000 & 100 & 160 & 20 & 55 & 140 & 25 & 157 & 125 & 50 & 86 & 80 & 60 \\
4 & - & - & - & 286 & 191 & 30 & 130 & 149 & 38 & 185 & 266 & 50 & 86 & 90 & 60 \\
5 & - & - & - & 70 & 70 & 100 & 90 & 150 & 250 & 170 & 217 & 75 & 26 & 90 & 100 \\
6 & - & - & - & 105 & 151 & 100 & 193 & 226 & 250 & 62 & 62 & 85 & 100 & 161 & 102 \\
7 & - & - & - & 60 & 250 & 666 & 119 & 220 & 477 & 35 & 35 & 550 & 73 & 97 & 111 \\
8 & - & - & - & - & - & - & - & - & - & 58 & 53 & 1900 & 159 & 323 & 194 \\
9 & - & - & - & - & - & - & - & - & - & - & - & - & 113 & 277 & 315 \\
10 & - & - & - & - & - & - & - & - & - & - & - & - & 27 & 18 & 5000 \\
\hline
\end{tabular}\end{center}
\end{table}
We created a baseline by using a depth-first selection rule for traversing the 
branch-and-price tree, as outlined in \cite{daSilva:24} and compared it with the 
heuristic selection strategy. The maximum runtime was set to two 
hours\footnote{Test system: Ubuntu 24.04.2, Gurobi 11.0.3, AMD Ryzen 5800X CPU, 64GB RAM}.
Unlike the branch-and-price approach, the MIP in \cite{or2024} could not find any feasible 
solutions within the time limit.
Table \ref{tab:results} demonstrates that the heuristic selection rule significantly 
reduced the number of used patterns compared to the depth-first selection rule. 
However, for problems involving ten or more item types and high item multiplicities, 
the depth-first selection method found solutions using fewer bins but more patterns 
within the time limit.
\begin{table}[tb]
\caption{Results: comparing selection strategies}
\label{tab:results}
\scriptsize{\begin{center}
\begin{tabular}{l@{\hspace{0.5cm}}r@{\hspace{0.25cm}}r@{\hspace{0.25cm}}r@{\hspace{0.25cm}}r@{\hspace{0.25cm}}cr@{\hspace{0.25cm}}r@{\hspace{0.25cm}}r@{\hspace{0.25cm}}r@{\hspace{0.25cm} }}\toprule
& \multicolumn{4}{c}{depth-first node selection} & & \multicolumn{4}{c}{heuristic node selection} \\
\cmidrule(lr){2-5} \cmidrule(lr){7-10}
record & patterns & bins & time & gap & & patterns & bins & time & gap \\
\midrule
$r_1$ & 13 & 19 & 1052 s & 0\% & & \textbf{4} & 19 & 1192 s & 0\%\\
$r_2$ & 27 & 57 & 7200 s & 3.1\% & & \textbf{6} & \textbf{55} & 1850 s & 0\%\\
$r_3$ & 33 & 108 & 7200 s & 1.8\% & & \textbf{11} & \textbf{103} & 1385 s & 0\%\\
$r_4$ & 46 & 54 & 7200 s & 6.4\% & & \textbf{3} & 54 & 7200 s & 7,1\%\\
$r_5$ & 87 & \textbf{112} & 7200 s & 13.2\% & & \textbf{8} & 192 & 7200 s & 49.3\%\\
\bottomrule
\end{tabular}\end{center}
}
\end{table}

The algorithm produces practical solutions. However, the difficulty in closing 
the MIP gap increases with the number of item types. This may be due to the fact 
that pricing is only solved heuristically, which can cause early termination of 
column-generation and suboptimal dual bounds.
To further improve the solutions, 
better pricing heuristics could be developed.
Adding 
conflict propagation, as described in \cite[p. 8]{daSilva:24}, may help to reduce the 
size of the search tree.

%%%%%%%%%%%%%%%%%%%%%%%%%%%%%%%%%%%%%%%%%%%%%%%%%%%%%%%%%%%%%%%%%%%%%%%%%%
%%%%%%%%%%%%%%%%%%%%%%%%%%%%%%%%%%%%%%%%%%%%%%%%%%%%%%%%%%%%%%%%%%%%%%%%%%
\bibliographystyle{plain}
\bibliography{ax_or2025}

\end{document}